# Decarbonization analysis on residential end uses in the emerging economies

Ran Yan[1], Minda Ma[2,3*]

1 School of Management Science and Real Estate, Chongqing University, Chongqing, 400045, PR China

2 Building Technology & Urban Systems Division, Lawrence Berkeley National Laboratory, Berkeley, CA 94720, United States

3 School of Architecture and Urban Planning, Chongqing University, Chongqing, 400045, PR China

(Corresponding Author, maminda@lbl.gov; minda.ma@cqu.edu.cn)

**ABSTRACT**

This study explores the historical emission patterns and decarbonization efforts of China and India, the largest emerging emitters in residential building operations. Using a novel carbon intensity model and structural decomposition approach, it assesses the operational decarbonization progress over the past two decades. Results show significant decarbonization, with China and India collectively reducing 1498.3 and 399.7 MtCO$_2$, respectively. Electrification notably contributed to decarbonizing space cooling and appliances in both countries.

## NONMENCLATURE

| | |
|---|---|
| *Abbreviations* | |
| DSD | Decomposing structural decomposition |
| HCE | Household consumption expenditure |
| *Symbols* | |
| $C$ | Carbon emissions |
| $c$ | Carbon intensity |
| $E$ | Energy consumption |
| $e$ | HCE-related energy intensity |
| $G$ | GDP |
| $g$ | GDP per capita |
| $H$ | Family households |
| $k$ | The emission factors |
| $P$ | Population size |
| $p$ | Household size |
| $S$ | HCE |
| $s$ | Household expenditure index |
| $w$ | End-use energy structure |

## 1. INTRODUCTION

The building sector contributes 27% of global carbon emissions, with residential emissions reaching a record high of 60% after COVID-19. Given the significant rise in residential emissions in emerging economies, analyzing energy demand and decarbonization efforts of China and India, the major nations, is vital for shaping equitable carbon policies, particularly beneficial for other emerging economies in global climate negotiations.

Aside from macroeconomic indicators, discussion on the impact factors of operational carbon emissions from residential buildings should investigate various end-use energy consumption in households [1], directly affected by user behavior [2] and facility technology [3]. Current studies rarely pursue a comparison of decarbonization changes in residential buildings affected by end-use performances between China and India across the 21st century [4]. Therefore, the following questions should be addressed to bridge this gap and evaluate the future decarbonization potential of residential buildings in China and India, seeking a fair emissions cap:

• What are the historical processes of operational emissions and the corresponding decarbonization?

• What is the impact of end-use performances on operational decarbonization?

To address these questions, this study evaluates the operational decarbonization progress of residential buildings in China and India in the 21st century via decomposing structural decomposition (DSD) approach. Specifically, considering economic, societal, and behavioral aspects, an emission model featuring end-use consumption is developed for identifying factors affecting carbon intensity changes, and the impact of various end uses on the carbon intensity is further investigated. Furthermore, six scales of decarbonization are utilized to evaluate the historical processes of decarbonizing residential buildings.

The main novelty lies in introducing an efficient data-driven model to analyze the decarbonization patterns of residential building operations, particularly in emerging economies like China and India, over the past two decades. Achieving deep decarbonization early in major emitters such as India and China can unlock additional carbon budget for other emerging economies, aiding progress towards the 1.5°C goal.

## 2. LITERATURE REVIEW

In previous studies, multiple perspectives have been employed to analyze the impact of end uses on energy and emissions of residential building operations in China [5,6]. For example, Zheng et.al [4] conducted a household survey to provide insight into the characteristics of end-use energy consumption for a specific year. Some researchers have explored carbon emissions from particular end-use performances in detail [7]. Swan et al. [8] reviewed residential energy modeling approaches to further compare end-use energy consumption between China and other countries. In terms of related studies on India, due to the widespread use of dirty energy sources such as biomass and kerosene, a low electrification rate, and the slow pace of urbanization, researchers have primarily focused on energy consumption rather than carbon emissions when discussing energy transitions in Indian households [9] and energy efficiency improvements for home appliances. Although the India National Sample Survey Organization periodically released detailed information on household energy consumption based on a national household sample survey, a time lag and data incoherence have resulted in a scarcity of recent studies related to end-use carbon emissions of residential building operations [10]. Projections of future household energy consumption patterns suggested that India's carbon emissions from residential building operations in 2050 may increase tenfold compared to 2005 levels [11]. Consequently, it is worth examining and comparing the decarbonization efforts in residential buildings in the top emerging emitters.

Although there are some related studies on global or individual scales, there is a lack of in-depth analyses focusing on the competition between China and India [12]. Furthermore, due to data acquisition challenges, short-term or outdated studies failed to accurately depict the end-use carbon emissions and decarbonization changes in residential building operations influenced by lifestyle changes, technology advancements, economic recession, and the COVID-19. A comprehensive comparative analysis at the national level since the millennium is crucial for evaluating the future potential for decarbonization and seeking fair emission allowances.

## 3. METHODS AND MATERIALS

*3.1 Emission model of residential building operations*

End-use energy consumption includes primary energy consumption, retail electricity, and heating. Residential households utilize operational energy through various end uses. This study decomposes residential building end uses into six types: space cooling, space heating, lighting, water heating, cooking, and appliances with others. Carbon emissions result from these activities, as represented in Eq. (1).

$$C = \sum_{i=1}^{6} C_i \quad (1)$$

The carbon emission intensity, measured as carbon emissions per household, is influenced by six factors: population, households, GDP, household consumption expenditure (HCE), energy consumption, and carbon emissions [13]. These factors are interrelated and represented by the following identity:

$$c = \frac{C}{H} = \frac{E}{S} \cdot \frac{C}{E} \cdot \frac{P}{H} \cdot \frac{G}{P} \cdot \frac{S}{G} \quad (2)$$

Combining Eqs. (1)-(2), the carbon emission intensity associated with end use $i$ in residential building operations can be written as:

$$c_i = \frac{C_i}{H} = \frac{E_i}{S} \cdot \frac{C_i}{E_i} \cdot \frac{P}{H} \cdot \frac{G}{P} \cdot \frac{S}{G} = e_i \cdot k_i \cdot p \cdot g \cdot s \quad (3)$$

*3.2 DSD-based decomposition of carbon intensity*

By applying the DSD approach to the emission model described in Section 3.1, this study was able to decompose the operational carbon intensity of residential building in India and China from 2000 to 2020 and explore the various factors contributing to carbon intensity changes.

The DSD method involves introducing $e$ as the sum of the household expenditure-related energy intensity, which is defined as $e = \sum_{j=1}^{6} e_i$. Additionally, $w_i$ is defined as the share of $E_i$ in $E$, or end-use energy structure, which can also be expressed as $w_i = \frac{E_i}{E}$. Eq. (5) was established as:

$$c = \sum_{i=1}^{6} e \cdot k_i \cdot w_i \cdot p \cdot g \cdot s \quad (5)$$

Then, Eq. (6) is required to derive the total differential. The shift component $dF_i$ and slack component $dF$ are introduced to express the change in $w_i$ [1]. The equations can be yielded in Eq. (7).

For space considerations, the matrix form expansion of Eq. (7) is omitted here. Given that the DSD method is based on Euler's method of numerical integration, this study partitioned the changes in exogenous variables into N segments, ensuring that N is sufficiently large to make each segment small but not infinitesimal. This approach allows for a more approximate estimation of the cumulative impact of exogenous variables on endogenous variables. In accordance with the aforementioned principle, the matrix form expansion of Eq. (7) is condensed into a generalized expression denoted as Eq. (8). Subsequently, the variations in each segment are represented by Eq. (9).

$$\begin{cases} Dc = \sum_{i=1}^{6}\left(\begin{array}{c}\frac{\partial c_i}{\partial e}de + \frac{\partial c_i}{\partial p}dp + \frac{\partial c_i}{\partial g}dg + \frac{\partial c_i}{\partial s}ds + \\ \frac{\partial c_i}{\partial k_i}dk_i + \frac{\partial c_i}{\partial w_i}dw_i\end{array}\right) \\ dw_i = dF_i + dF \\ \sum_{i=1}^{6} dw_i = 0 \end{cases} \quad (7)$$

$$\boldsymbol{A} \cdot d\boldsymbol{y} = \boldsymbol{B} \cdot d\boldsymbol{z} \quad (8)$$

$$\begin{cases} \boldsymbol{D}^{(n)} = \left(\boldsymbol{A}^{(n-1)}\right)^{-1} \cdot \boldsymbol{B}^{(n-1)} \cdot diag(d\boldsymbol{z}) \\ d\boldsymbol{y}^{(n)} = \boldsymbol{D}^{(n)} \cdot \boldsymbol{j} \\ \boldsymbol{z}^{(n)} = \boldsymbol{z}^{(n-1)} + d\boldsymbol{z} \\ \boldsymbol{y}^{(n)} = \boldsymbol{y}^{(n-1)} + d\boldsymbol{y}^{(n)} \\ \boldsymbol{A}^{(n)} = f(\boldsymbol{z}^{(n)}, \boldsymbol{y}^{(n)}) \\ \boldsymbol{B}^{(n)} = g(\boldsymbol{z}^{(n)}, \boldsymbol{y}^{(n)}) \end{cases} \quad (9)$$

where $d\boldsymbol{z} = diag(d\boldsymbol{z}) \cdot \boldsymbol{j} = \frac{\Delta \boldsymbol{z}}{N}$, $diag(d\boldsymbol{z})$ is the diagonal matrix composed of the differential element $d\boldsymbol{y}$, and $\boldsymbol{j}$ is a vector consisting of ones. For $n = 1, 2, \cdots, N$, this study set N=16000 to ensure adequate precision, as stipulated in the original research article on the DSD method [1].

The desired decomposition is achieved through iterative computations, with contributions from each change being summed:

$$\boldsymbol{D} = \sum_{n=1}^{N} \boldsymbol{D}^{(n)} \quad (10)$$

In this study, the method described above can be used to measure the impact of changes in exogenous variables on carbon intensity. Using the DSD method, the changes in carbon intensity can be decomposed as:

$$\Delta c|_{0 \to T} = \Delta e_{\text{DSD}} + \Delta p_{\text{DSD}} + \Delta g_{\text{DSD}} + \Delta s_{\text{DSD}} + \Delta k_{\text{DSD}} + \Delta w_{\text{DSD}} \quad (11)$$

where $\Delta k_{\text{DSD}}$ and $\Delta w_{\text{DSD}}$ can be further decomposed, which yields the following:

$$\begin{cases} \Delta k_{\text{DSD}} = \sum_{i=1}^{6} \Delta k_i \\ \Delta w_{\text{DSD}} = \sum_{i=1}^{6} \Delta w_i \end{cases} \quad (12)$$

## 4. RESULTS

### 4.1. Carbon intensity changes in residential building operations

The overall carbon intensity has increased for both countries over the last two decades, with a higher average growth rate of 2.5% per year in India and 1.4% per year in China. By dividing the decades into four stages of five years each, the performance of each stage between China and India dynamically changed according to the technical, socioeconomic, and related strategies, as shown in Fig.1.

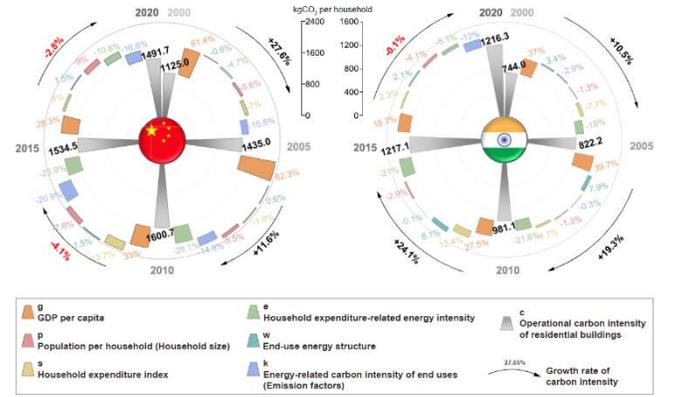

**Fig. 1.** *Operational carbon intensity changes in residential buildings in China and India (2000–2020).*

In China, the operational carbon intensity experienced a gradual slowdown in growth rate between 2000 and 2010 (from 1125 to 1601 kgCO$_2$ per household), reaching an annual peak in 2012 at 1606.4 kgCO$_2$ per household. Since 2010, the Chinese government has further strengthened its decarbonization strategies. Consequently, the operational carbon intensity changes manifested as continuous and slight declines between 2010 and 2020, with an average of -0.9% per year.

In India, the operational carbon intensity of residential buildings demonstrated sustained growth between 2000 and 2015 (from 744 to 1217 kgCO$_2$ per household). Notably, the operational carbon intensity in India reached its annual peak at 1283.3 kgCO$_2$ per household in 2018, before decreasing by -2.6% each year in 2019 and 2020. This decline was due to India's economic contraction amid the COVID-19 pandemic, resulting in restrictions on fossil fuel consumption.

Further analysis of carbon intensity drivers highlighted GDP per capita as the predominant factor, consistently showing a positive impact. From 2000 to 2020, the total contribution of GDP per capita in China was 226.4%, while in India, it was 144.5%. Conversely, household expenditure-related energy intensity had the most significant negative impact on carbon intensity reduction in both countries, with China at -90.6% and India at -78.1%. This was followed by emission factors, which contributed a negative effect of -87.4% in China and -23.0% in India. The role of household expenditure-related energy intensity in promoting decarbonization is evident in redirecting energy consumption towards cleaner sources, as seen in China's transition to cleaner heating fuels [15-16]. Moreover, the increased adoption of renewable energy in household electricity usage presents a technical avenue for enhancing the decarbonization potential of emission factors [17-19].

*4.2. Impact of end uses on the operational carbon intensity changes*

The study also examined the impact of various end-use behaviors on operational carbon intensity. It divided the contribution of emission factors, represented as pale blue fans in Fig. 1, into six different end uses, as illustrated in Fig. 2. The fan charts in Fig. 2a and 2c depict the decomposition of contribution rates of emission factors across different end uses over four stages from 2000 to 2020 for India and China, respectively. Fig. 2b presents bar charts that transform the contribution rates into the absolute value of carbon intensity changes, allowing for a more intuitive analysis.

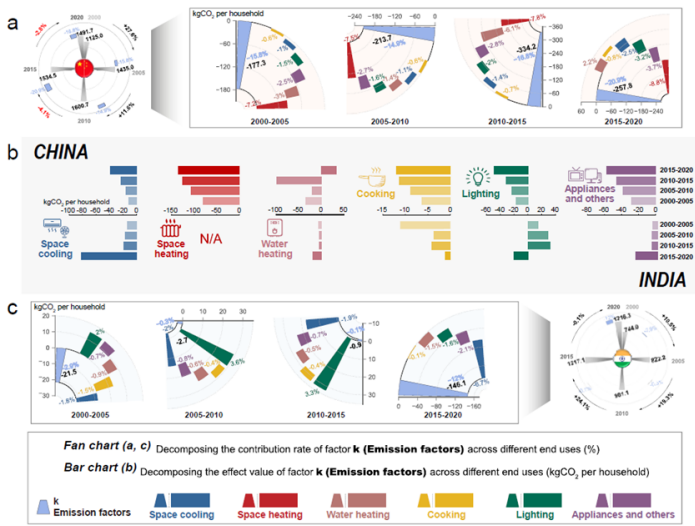

**Fig. 2.** Emission factor effects of various end uses on carbon intensity changes in residential buildings in (a) China and (c) India from 2000 to 2020; (b) absolute value of the emission factor effects on carbon intensity changes among various end uses (2000-2020).

In general, In China, each end use's emission factor negatively impacted operational carbon intensity, contributing positively to decarbonization. From 2000 to 2020, these negative effects gradually increased, reflecting China's evolving energy structure. Over the past two decades, this evolution included increased electrification of residential buildings (the electrification rate rose from 4.5% in 2000 to 27.2% in 2020) and greater adoption of natural gas in urban areas (with users increasing from 1.8% in 2000 to 29.3% in 2020).

However, unlike China, not all emission factor effects in India negatively impacted operational carbon intensity. While space cooling, appliances, and other end uses consistently reduced carbon intensity from 2000 to 2020, lighting's contribution shifted from positive to negative between 2015 and 2020. End uses related to electricity consumption notably drove down carbon intensity in India, particularly from 2015 to 2020, confirming the improvement in India's residential building electrification rate, which increased from 5.4% in 2000 to 17.8% in 2020.

Specifically, in China, space heating had the most significant negative impact on carbon intensity reduction (-39.8%), followed closely by appliances and other end uses (-15.1%). Lighting, water heating, space cooling, and cooking contributed -10.7%, -10.6%, -7.8%, and -3.4%, respectively. In India, space cooling was the leading negative contributor (-17.5%), followed by appliances and other end uses (-5.8%). Lighting shifted from positive to negative between 2015 and 2020 but remained the most positive contributor overall (7.6%). Water heating and cooking contributed -4.6% and -2.7%, respectively. Space heating wasn't considered due to minimal demand in India's tropical climate.

Overall, emission factor effects efficiently drove residential building decarbonization in top emerging economies like China and India. Electrification notably enhanced contributions from space cooling, appliances, and lighting to decarbonization. Energy optimization in residential space heating in China also yielded significant decarbonization benefits. However, challenges persist in systematizing and purifying household energy consumption for water heating and cooking in underdeveloped areas, hindering operational decarbonization in both countries. Additionally, changes in end-use energy structures influenced operational carbon intensity changes in residential buildings, as revealed by the DSD method's structural change effects, though these effects contributed minimally to carbon intensity changes.

## 5. DISCUSSION

This study examined the historical decarbonization performance of residential buildings in India and China using six decarbonization scales. Fig. 3 shows the total decarbonization and efficiency trends from 2000 to 2020. In China, cumulative decarbonization reached 1498.3 MtCO$_2$, with a 11.5% efficiency over 20 years. Phased accumulation shares were 17%, 24%, 29%, and 30%, with efficiency ranging from 11% to 12%, indicating a steady 5.7% annual increase. India's cumulative decarbonization was 399.7 MtCO$_2$, with a 7.5% efficiency. Phased shares were 16%, 21%, 23%, and 40%, with a 7.3% yearly growth since 2015, peaking at 13.6% in 2020. However, neither country has reached a significant annual decarbonization peak for residential operations, signaling the need for improvement. With decarbonization slowing in developed nations, China and India face rising pressure to meet the 1.5°C target.

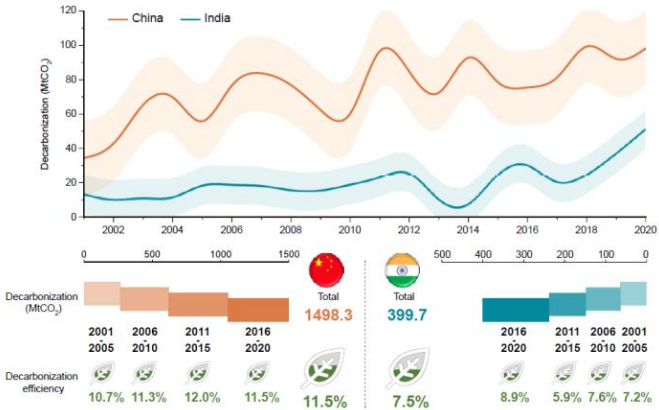

**Fig. 3.** *The total decarbonization and decarbonization efficiency of residential building operations in India and China in 2000-2020.*

In addition, this study aimed to analyze the decarbonization of residential building operations in India and China across various metrics, including decarbonization intensity, per capita, per floor area, and per household expenditure (Fig. 4a). In China, decarbonization intensity averaged 167.3 kgCO$_2$ per household per year, growing at 3.7% annually. Despite yearly fluctuations, it stabilized above 150.0 kgCO$_2$ per household after peaking in 2011 at 255.2 kgCO$_2$ per household. From 2015 to 2020, China's share in total cumulative decarbonization intensity reached 26%. Conversely, India's decarbonization intensity surged, particularly post-2015, with its share reaching 36% from 2015 to 2020. In 2020, India's intensity nearly matched China's at 164.8 kgCO$_2$ per household, reflecting its robust 7.3% yearly growth.

In terms of decarbonization per capita (Fig. 4b), China has significantly outpaced India, with an average annual increase of 55.5 kgCO$_2$ per capita compared to India's 16.5 kgCO$_2$ per capita. China's growth rate stood at 5.1%, slightly lower than India's 5.8% over the past two decades. Between 2015 and 2020, the share of the accumulation in the total cumulative decarbonization per capita in India reached 37%, coinciding with the rapid development of decarbonization efforts in India.

Decarbonization per floor area (Fig. 4c) has shown relatively similar trends between India and China. China's average was 1.7 kgCO$_2$ per square meter per year, with a 1.9% annual increase. India's average was slightly lower at 1.5 kgCO$_2$/m$^2$/year, growing at 3.0% annually. India's per floor area decarbonization reached its peak in 2020 at 2.7 kgCO$_2$/m$^2$, higher than China's 1.6 kgCO$_2$/m$^2$, and is expected to continue increasing significantly.

Decarbonization per household expenditure (Fig. 4d) in China sharply declined since 2004, with an average annual reduction of -4.3%, driven by rapid spending power growth outpacing decarbonization progress. Conversely, India saw a slight decrease of -0.01% per year over the same period. Values converged around 2015, reaching 10.7 kgCO$_2$ per thousand dollars in China and 9.0 kgCO$_2$ per thousand dollars in India in 2020. Similar to the differences in decarbonization intensity between the two countries, a reversal can also be anticipated in the coming years.

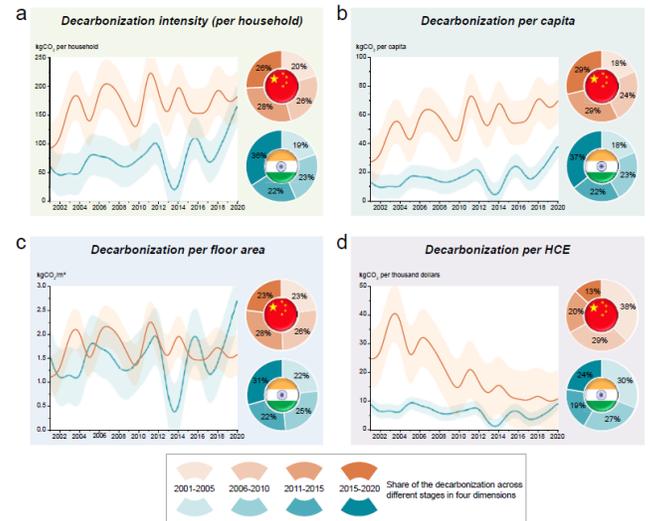

**Fig. 4.** *The historical performance of (a) decarbonization per household, (b) per capita, (c) per floor area, and (d) per household expenditure of residential building operations in India and China during 2000-2020.*

## 6. CONCLUSIONS

This study assessed operational decarbonization in residential buildings in India and China during the 21st century using DSD method. It developed a carbon intensity model to identify factors affecting carbon changes and investigated the impacts of various end uses on carbon intensity. Six decarbonization scales were employed to track historical processes. These efforts can significantly shape a fairer carbon budget framework in global climate diplomacy, particularly benefiting India and China. Key findings are summarized below.

- **From 2000 to 2020, operational carbon intensity increased by 1.4% annually in China (from 1125 to 1492 kgCO$_2$ per household) and by 2.5% annually in India (from 744 to 1216 kgCO$_2$ per household).** China's residential buildings reached a peak intensity of 1606.4 kgCO$_2$ per household in 2012, while India peaked at 1283.3 kgCO$_2$ per household in 2018. GDP per capita made the most significant positive contribution, totaling 226.4% in China and 144.5% in India. Conversely, household expenditure-related energy intensity was the most significant negative contributor (-90.6% in China and -78.1% in India), followed by emission factors (-87.4% in China and -23.0% in India), both crucial for decarbonizing residential building operations.

- **Building electrification promoted the end-uses' emission factor effects on decarbonization (e.g., space cooling contributed -87.7 and -130.2 kgCO$_2$ per household in China and India, respectively).** In China, space heating was the most significant positive contributor to decarbonization, reducing carbon intensity by -39.8% and -447.9 kgCO$_2$ per household from 2000 to 2020. Appliances and other end uses followed with contributions of -15.1% (-169.7 kgCO$_2$ per household), while lighting contributed -10.7% (-120.6 kgCO$_2$ per household). In India, space cooling was the primary positive contributor, with a reduction of -5.8% (-43.4 kgCO$_2$ per household), followed by appliances and others. Despite lighting's positive effect from 2015, it remained the most significant negative contributor (-7.6%, 56.4 kgCO$_2$ per household).

- **China and India collectively decarbonized 1498.3 and 399.7 MtCO$_2$ of residential building operations from 2000 to 2020, without reaching an annual decarbonization peak.** China led with a decarbonization efficiency of 11.5%, compared to India's 7.5%, while the United States had an efficiency of 8.5%, revealing an inverted-U shaped relationship between decarbonization efficiency and national development. India's operational decarbonization intensity is expected to surpass China's in the coming years, as India reached 164.8 kgCO$_2$ per household in 2020, close to China's 182.5 kgCO$_2$ per household. Decarbonization per floor area and per household expenditure have been similar between China and India in recent years. China should focus on deep building decarbonization to free up emission allowances for construction industry development in emerging economies, while India should balance energy decarbonization with urbanization-driven economic growth for a significant low-carbon transition.